# Features of the Contact Angle Hysteresis at the Nanoscale: A Molecular Dynamics Insight

Viktor Mandrolko[1], Guillaume Castanet[1], Sergii Burian[2], Yaroslav Grosu[3,4], Liudmyla Klochko[1*], David Lacroix[1], and Mykola Isaiev[1]

[1]*Université de Lorraine, CNRS, LEMTA, 54000 Nancy, France*
[2]*Faculty of Physics, Taras Shevchenko National University of Kyiv, 64 Volodymyrska Street, Kyiv 01601, Ukraine*
[3]*Centre for Cooperative Research on Alternative Energies (CIC energiGUNE), Basque Research and Technology Alliance (BRTA), Alava Technology Park, Albert Einstein 48, 01510 Vitoria-Gasteiz, Spain*
[4]*Institute of Chemistry, University of Silesia, Szkolna 9 street, 40-006 Katowice, Poland*
*current affiliation Université de Lorraine, CNRS, Inria, LORIA, 54000 Nancy, France*

Corresponding authors: Viktor Mandrolko, email: viktor.mandrolko@univ-lorraine.fr; Mykola Isaiev, email: mykola.isaiev@univ-lorraine.fr

## Abstract

Understanding the physics of three-phase contact line between gas, liquid, and solid is important for numerous applications. At the macroscale, the three-phase contact line response to an external force action is often characterized by a contact angle hysteresis, and several models are presented in the literature for its description. Yet, there is still a need for more information about such model applications at the nanoscale. In this study, a molecular dynamics approach was used to investigate the shape of a liquid droplet under an external force for different wetting regimes. In addition, an analytic model for describing the droplet shape was developed. It gives us the possibility to evaluate the receding and advancing wetting angle accurately. With our modelling, we found that the interplay between capillary forces and viscous forces is crucial to characterize the droplet shape at the nanoscale. In this frame, the importance of the rolling movement of the interface between liquid and vapor was pointed out. We also demonstrate that in the range of the external forces when capillary forces are most significant compared to others, hysteresis is well described by the macroscale Cox-Voinov model.

## Introduction

Surface wettability is of paramount importance for a wide range of natural and technological processes spanning from the mystery of the origin of life [1] to wastewater remediation [2], biomedical applications [3], catalysis [4], heat and mass transfer [5–7], etc. This issue has been considered for decades at macroscale, but, given the rapid development of nanomaterials, where surfaces and interfaces play a dominant role, questions about the wettability of nanostructures become more important [5–8]. For example, the wettability of nanoporous materials such as zeolites, metal–organic frameworks (MOFs), covalent organic frameworks (COFs), and silica gels is a key issue for catalysis, liquid/liquid separation, chromatography, energy storage and conversion, and other processes [9–13]. However, going behind those materials and applications where small scales matter entails to develop solid and accurate models to understand the physic at play.

At the macroscale, the interactions between a solid and a liquid can be described in terms of wettability. This parameter characterizes a liquid ability to spread over a solid surface. Such behaviour results from the balance between forces of adhesion and cohesion. The most convenient way to quantify the "wetting state" is given by the "contact angle" (CA also noted such as $\theta$ hereafter). At the first sight, CA is a very clear and easily identifiable parameter that can be inferred from the slope of a line tangent to the liquid interface at the contact point of three phases: liquid, solid, and vapor. According to the Young's equation, CA depends on the interfacial free energy of the solid-liquid $\gamma_{sl}$, solid-vapour $\gamma_s$, and liquid-vapour $\gamma_l$ interface:

$$cos\,\theta = \frac{\gamma_s - \gamma_{sl}}{\gamma_l} \qquad (1)$$

The main disadvantage of this formulation is that it is only valid for rigid, flat, smooth and homogenous surfaces without any external forces acting on the fluid. Those are virtually impossible to obtain in "real life" applications. This point leads to several issues, especially to find the equilibrium CA in real systems. When addressing practical cases, there is a whole range of the local Gibbs free energy minima due to the surface imperfections such as roughness and heterogeneities, which results in the existence of a spectrum of possible contact angles. The maximum and minimum CA of this spectrum is referred as the "advancing" and "receding" angles. Those angles can be reached by increasing/decreasing droplet's volume and measuring the CA right before the contact line moving.

The situation becomes even more complex when external forces are interacting with the fluid molecules. These forces can stem from various sources, including for example: pressure differences in a porous media [14], magnetic fields [15], electroosmotic flow from electrical forces [16], and mechanical forces such as those exerted by an AFM [17]. Another case is high velocity fluid impinging a wall. For the latter, the kinetic energy of the droplet before impact is partially transformed into surface energy as the droplet is spreading, and the contact angle varies with the impact velocity.

In all these examples, the equilibrium condition at the three-phase contact line depends on the applied force which leads to a change of the wetting angle. There is, therefore, a noticeable difference between the equilibrium or the static contact angle predicted by Eq. 1 and the dynamic contact angle. Besides, with the applied external force the CA is changing along the droplets contact line depending on the direction of the force. The angles formed at the front and back of a droplet right before the movement are the maximum and minimum CA. Even though, strictly speaking, they do not align the exact definition of the advancing and receding contact angles due to the break of axial symmetry [18] based on its closeness in values and their common usage in the field [19–21] we choose to stick to the terms "advancing and receding contact angles".

The range CA values that can be taken onto a given surface is referred as the contact angle hysteresis (CAH). From the above description, there are many ways for a CAH to occur. It can be observed in situations where a contact line pinning is present [22] (static CAH), and situations where the contact line is moving (dynamic CAH) [21]. The value of CAH is strongly connected with the lateral adhesion force required to move a drop across a surface, which is expressed as:

$$F_c = \gamma_l \, r \, k \, (\cos\theta_a - \cos\theta_r) \tag{2}$$

where $r$ is the distance between the two points of contact with the solid surface, $\theta_a$ and $\theta_r$ – advancing and receding contact angles. $k$ is an empirical factor that depends on the geometry of the droplet [23]. Eq. 2 demonstrates that without contact angle hysteresis, drops would slide down even at a minimal inclination angle of the substrate.

The static CAH is essential in many processes, such as surface coating with liquid, evaporation [27,28], condensation [29], boiling [30] and heat transfer processes [31]. It is critical in various applications: printing and painting (contact dispensing and inkjet printing), soldering, percolation, distributing of herbicides and insecticides. Static CAH is an important characteristic that depends on both properties of the liquid and the solid surfaces.

While the measurable apparent CA is significantly influenced by the surface heterogeneity and the contact line pinning, CAH is less impacted, making it a more reliable representation of the interaction between the liquid and the solid surfaces [24]. Moreover, some methods estimate the value of thermodynamically equilibrium CA based on CAH [25]. Besides that, in a study [26], it was also demonstrated that Young's static equilibrium CA solely governs the CAH on a smooth and chemically homogeneous surface.

While the static CAH varies in a specific range with the surface properties, the dynamic one is expected to be defined by the motion velocity of the contact line (or a point) [20,27]. Furthermore, since wetting properties caused by different factors like hydrodynamic dissipation [28], surface adaptation [29] and energy barriers [30,31], the advancing and receding contact angles also depend on them.

In the macrosystems, the interphase transition region between two adjacent homogeneous subsystems is considered to have zero-thickness. While approaching the nanoscale interfacial region started to be more and more important. Specifically, one should take into account that the transition between adjacent homogeneous thermodynamic phases happens smoothly in some volume with finite thickness [32]. When the characteristic dimensions of a nanoscale system are comparable to the thickness of this transition region, defining the interface that separates two phases can be challenging. Such definition is usually done based on iso-density surface, although its "numerical treatment" varies according to different works [33–36]. In addition to that, two distinct surfaces are usually considered: the surface of tension and the equimolar surface. Both may be used as interfaces separating different phases. The location of these interfaces is different and the well-known "Tolman length" $\delta$ represents the distance between these two surfaces. Based on this parameter it is possible to assess the deviation from ideal behavior at the interface, so it plays a crucial role in understanding the thermodynamic properties of nano-scale systems. The change in surface tension observed in nanoscale droplets can be calculated based on the Tolman equation:

$$\gamma_l = \frac{\gamma^\infty}{\left(1 + \frac{\delta}{R_c}\right)} \quad (3)$$

where $\gamma^\infty$ is the surface tension of the flat surface of liquid and $R_c$ the curvature radius of the droplet. It should be noted that the sign and value of the Tolman length for each individual liquid can vary with the temperature and the droplet radius [37]. For example, the dependence of water's surface tension on droplet size becomes noticeable for spherical droplets when the radius of curvature falls below 100 Å [38].

Furthermore, in the vicinity of the three-phase contact line, each phase affects the interaction between others, leading to "liquid's shape bending" [39]. This effect is usually considered by introducing the line tension $\tau$, which is the additional free energy per unit length present at a contact line where three distinct phases that coexist. [40]. Taking it into account, the Young's equation can be presented as follows:

$$\cos\theta = \frac{\gamma_s - \gamma_{sl}}{\gamma_l} - \frac{\tau/\gamma_l}{R_c} \quad (4)$$

Here, the second term of Eq. 4 depends on the curvature of the three-phase contact line, and thus vanishes for a macroscale droplet. The value of line tension lies in the range of $10^{-12}$ — $10^{-10}$ N [40] and, knowing the surface tension, one can estimate the scale on which the line tension plays an important role as:

$$\xi = \frac{\gamma_l}{\tau} \quad (5)$$

Where ξ is called line tension length. It is worth noting that the contact line of a cylinder is less pronounced as the curvature of the contacting line is close to zero [41]. Besides, due to the significant surface-to-volume ratio in a nanoscale system, adsorption makes a noticeable contribution to the wettability [42].

Therefore, despite the numerous studies in this area, additional research must be provided to understand the overall phenomena thoroughly. For example, CAH is experimentally observed for smooth, homogeneous surfaces [43] and even on free homogeneous liquid films [44,45], which contradicts the recent molecular dynamics studies that showed that the CAH effect stems from surface roughness if it is homogenous [46]. In addition to the factors contributing to the wetting angle's hysteresis, accurate measurement of this angle at the nanoscale remains challenging. Currently, the most used method involves approximation of the droplet profile with circles and determining its intersection angle with the solid surface [47]. Given the limited resolution provided by the molecular dynamics method under reasonable calculation time, it is difficult to determine the circle that describes the profile close enough to the contact line in the case of the non-spherical/non-circular droplet shape. At the same time, in the general case, it would be incorrect to approximate the shape of the droplet under the external forces by two circles representing the droplet's receding and advancing halves.

In this study, we present a simple model designed to predict the shape of a nanoscale water droplet moving under the influence of an external force. We employed molecular dynamics simulations to investigate the dynamic contact angle hysteresis of a nano-droplet of water. The applicability range of the model was evaluated by comparing the predicted droplet shape with those obtained through molecular dynamics simulations. It is demonstrated that at the nanoscale the droplet rolling speed is a key parameter to describe the CAH. This is in striking contrast to the macroscopic case where the linear velocity of the droplet is essential.

## 1. Simulation details

Classical molecular dynamics simulations were performed using the LAMMPS package [48,49]. The solid substrate was silicon-like structure with a surface crystal orientation plane (001). In the latter, atoms interact with the Stillinger-Weber potential [50]. A harmonic force was applied to the lower four layers of the silicon atoms to bind them to their initial positions. Such technique helps to avoid the displacement of the silicon substrate in the z-axis direction during the simulation. We should say that, in fact, we started from the Lennard-Jones wall as a substrate, but in that case no contact angle hysteresis was observed. The same is shown in this paper[46] : contact angle hysteresis arises due to surface roughness or heterogeneity. In the case of atomically smooth surface the spaces between the surface atoms create roughness leading to the hysteresis.

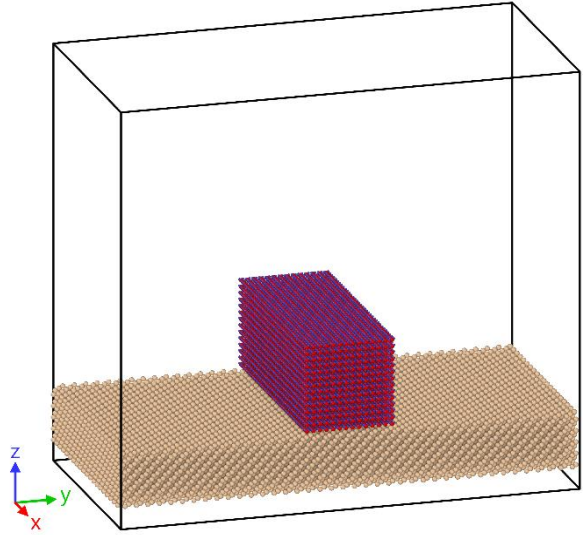

*Figure 1. Initial configuration of the considered droplet of water set on a silicon surface.*

For the liquid modelling the SPC/E water model was chosen [51]. In general case, one should take into account the interaction of silicon with the hydrogen [52]. However, following [46,53,54], the interaction between the substrate (silicon) and water was performed uniquely through the Lennard-Jones potential acting between oxygen and silicon atoms to be consistent with literature data:

$$V(r) = 4\varepsilon_{Si-O}\left(\left(\frac{\sigma_{Si-O}}{r}\right)^{12} - \left(\frac{\sigma_{Si-O}}{r}\right)^6\right) \quad (6)$$

where $r$ is the distance between atoms, $\varepsilon_{Si-O}$ is the depth of the potential well, $\sigma_{Si-O}$ is the distance at which potential is equal to zero. In order to create a broad range of wettability, the parameter $\varepsilon$ was varied in the range from 10 to 21 meV, while $\sigma_{Si-O}$ was maintained equal to 0.26305 nm which corresponds to a static CA range from 126 to 49°[41] on the flat silicon surface. For 10 meV the system is strongly "hydrophobic", while with 21 meV it is "hydrophilic".

To remove the complex spatial variation of CA along the three-phase contact line of a hemispherical droplet and at the same time reduce impact of droplet size-effect[38], the simulation of a "cylindrical droplet shape" on an atomically smooth surface in a system with periodic boundary conditions was chosen. The initial snapshot of the modelled system is shown in Figure 1. In the present case, only two contact angles are sufficient to describe this system.

Characteristic lengths of the simulation box and the number of atoms of each species (Si and $H_2O$) are given in Table 1.

Table 1. Lengths of the simulation box, silicon substrate, and water droplet, and the number of silicon wafer and droplet atoms.

| Simulation box | | 108.6 Å×271.5 Å×200.0 Å |
|---|---|---|
| Silicon substrate | | 108.6 Å×271.5 Å×43.44 Å |
| Initial configuration of a water droplet | | 108.6 Å×43.4 Å×43.40 Å |
| Number of atoms | Silicon | 4×20×50×8 = 32 000 |
| | Water | 3×35×14×14 = 20 580 |

The numerical simulation was conducted in two stages. I) During the first stage, the droplet was only subjected only to forces arising from its interaction with the substrate, with no external forces applied. In this case, after the initial configuration was created (as shown in Figure 1), all atoms in the system were given initial velocities corresponding to an average temperature of 1 K. Then, the system was gradually heated from 1 K to 300 K over a period of 0.3 nanoseconds and maintained at 300 K for an additional 5 ns period to ensure thermalization. The Nosé–Hoover thermostat in the canonical ensemble [55,56] was applied. One can see from (Figure 2), that such protocol gave a stable and equilibrated droplet with a cylindrical cap shape. The density profile was obtained by partitioning the simulation box into 1 Å thick bins along the Y and Z axes.

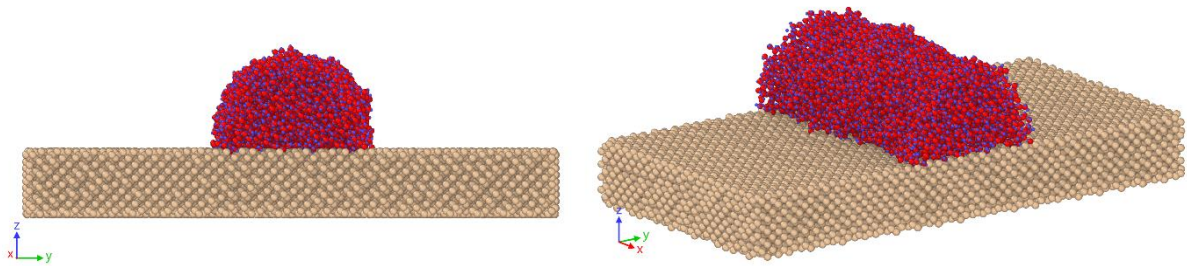

*Figure 2. Configuration of a droplet of water after thermalization.*

II) During the second stage, an external volumetric inertial-like force (with a constant acceleration) was set on each atom in the droplet along the Y-axis. The total force magnitude ranged from 0 to 1.386 nN, with an increment of 0.099 nN. As a result, the droplet first moved with some acceleration and then reached a uniform velocity mode within less than 1 ns. Subsequently, the particle trajectories were recorded for an additional 1 ns. It should be noted that the uniform velocity arises due to the balance between external force and the force of friction. The excess energy caused by the action of the non-conservative friction force was naturally dissipated by the thermostat applied to the system.

The fact that the droplet starts to move uniformly makes it possible to subtract the velocity of its centre of mass, eliminating all the movement except the internally created one. The uniform velocity of the centre of mass as a function of the applied external force is presented in Figure 3.

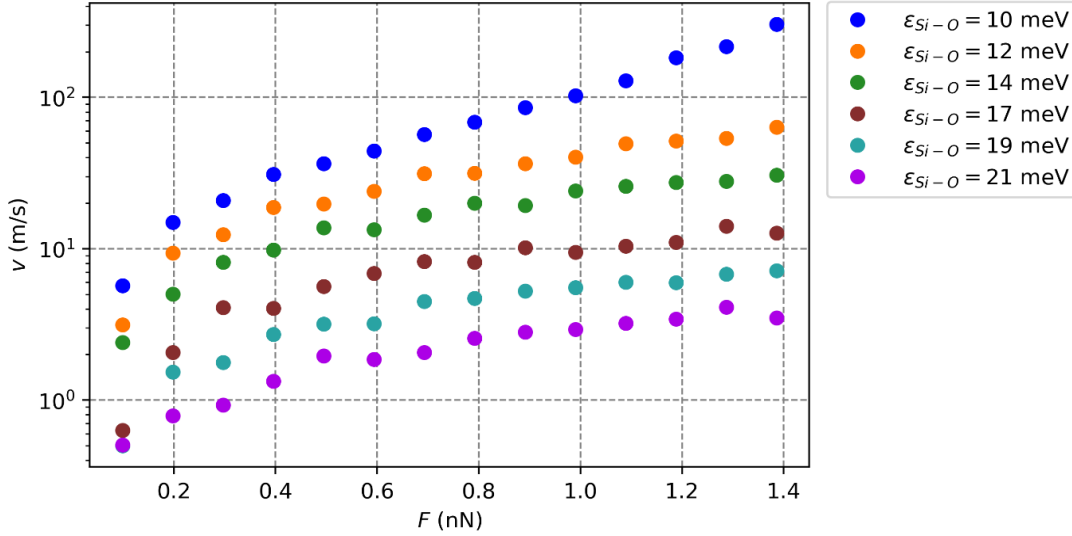

*Figure 3. Dependence of the droplet's velocity on the applied total external force.*

After velocity subtraction, based on 1000 snapshots of the atom's position with an interval of 1 ps between them, the density profile was averaged in the same way as for the stationary case. Hereafter, examples of the droplet shape, under the influence of different external forces and averaged density profiles are presented in Figure 4. The red line in Figure 4 is added to outline the edge of the droplet, defined as an equimolar surface corresponding to the half-density of bulk water.

For each value of $\varepsilon_{Si-O}$, a minimum of 15 simulations were performed, which corresponds to the number of considered applied external forces. For some values of $\varepsilon_{Si-O}$ (10, 15, 19, 21 meV), to make sure that the results are independent of the initial random distribution of temperatures, an additional 15 simulations were conducted. In addition, 15 more simulations were conducted to test the size effect. Each of these simulations cost an average of 4225 core-hours. In total, the calculations cost about 760,500 core-hours.

## 2. Model for the droplet shape

A simple model was also proposed to describe the shape of a droplet submitted to an external volume force. This model is valid for the case where liquid has no internal motion inside the droplet. In this case, the droplet shape results exclusively from a balance between the capillary and hydrostatic pressures. The geometrical configuration considered in the model is shown in Figure 5. It is a two-dimensional droplet on a plane surface under the influence of an external force. The fundamental equation of capillarity, which relates the hydrostatic pressure in the liquid droplet column to the radius of curvature of the droplet profile, is given by Eq. 7:

$$\Delta P = \gamma_l \frac{1}{R} \quad (7)$$

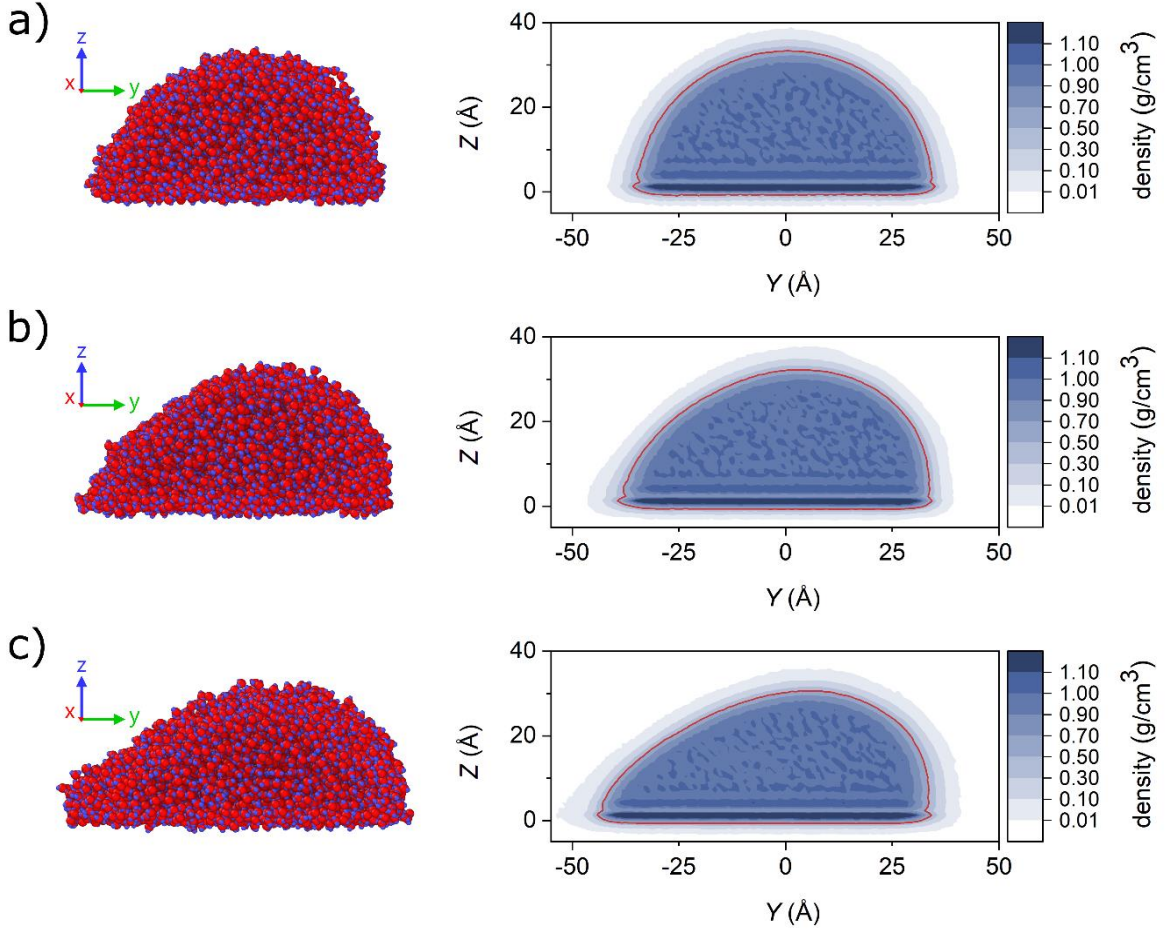

*Figure 4. Snapshots (left) and averaged density profiles (right) of the droplet under the influence of an external force of different extent: a) low (0.099 nN), b) medium (0.792 nN), c) high (1.386 nN) for the case of natural silicon-water wetting (ε = 15.75 meV)*

where $R$ is the radius of curvature of the droplet profile, $\Delta P$ is the hydrostatic pressure in the liquid droplet column at the considered point, $\gamma_l$ is the liquid-vapor surface tension. Denoting $h_c$ as the height of the liquid column, $\Delta \rho$ as the difference in density between the liquid and gas phases and $g$ as the inertia-like acceleration due to applied external force, $\Delta P$ can be written as:

$$\Delta P = \Delta \rho \, g \, h_c \quad (8)$$

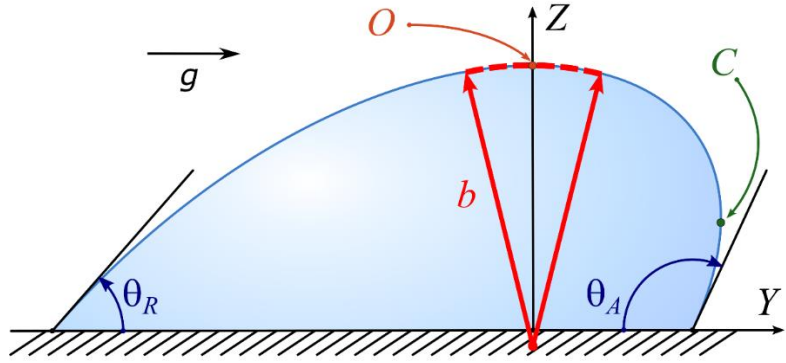

Figure 5. Geometrical configuration considered in the model.

The apex of the droplet (Point O in Figure 5) is chosen as a reference. The coordinate system is positioned such that the apex corresponds to $y = 0$. Introducing the radius of curvature at the apex $b$, the Young-Laplace equation (Eq. 7) an be rewritten as:

$$\gamma_l \frac{1}{R} = \Delta\rho\, g\, y + \frac{\gamma_l}{b} \qquad (9)$$

Thus at $y = 0$, $\Delta P = \gamma_l/b$ and at any other value of $y$, the change in $\Delta P$ is given by $\Delta\rho\, g\, y$. Eq. 9 can be rearranged to involve dimensionless parameters:

$$\frac{1}{R(y)/b} = Bo\, \frac{y}{b} + 1 \qquad (10)$$

where $Bo$ is the Bond number defined by:

$$Bo = \frac{\Delta\rho\, g\, b^2}{\gamma_l} = \frac{b^2}{a^2} \qquad (11)$$

In this expression, $a = \sqrt{\gamma/\Delta\rho g}$ corresponds to the capillary length. It should be noted that in this case the Bond number is not the classical one associated with gravity, but a more general one linked to the acceleration field generated by the external force. In our case, it represents the importance of external forces compared to surface tension forces.

The radius of curvature $R$ of a line $z(y)$ can be expressed from analytical geometry:

$$\frac{1}{R} = -\frac{z''}{\left(1 + z'^2\right)^{3/2}} \qquad (12)$$

Replacing $R$ by this expression in Eq. 10, the following is obtained:

$$\frac{-\tilde{z}''}{\left(1+\tilde{z}'^2\right)^{3/2}} = Bo\,\tilde{y} + 1 \qquad (13)$$

where $\tilde{z} = z/b$, $\tilde{y} = y/b$, $\tilde{z}' = d\tilde{z}/d\tilde{y}$ and $\tilde{z}'' = d^2\tilde{z}/d\tilde{y}^2$.

If a bend as at point C is present (which can also occur on the left side of a hydrophobic droplet), a numerical issue arises due to the tendency of $\tilde{z}'$ to infinity. To overcome it, it is necessary to consider the radius of curvature $R$ of the line $y(z)$ instead of $z(y)$, resulting in the transformation of Eq. 13 into:

$$-\frac{\tilde{y}''}{\left(1+\tilde{y}'^2\right)^{\frac{3}{2}}} = Bo\,\tilde{z} + 1 \qquad (14)$$

With boundary conditions:
At $\tilde{z} = z_C/b$ : $\tilde{y} = y_C/b$, y'=0
At $\tilde{z} = z_D/b$ : $\tilde{y} = y_D/b$, y'=0

Solving Eq. 13 and, if required, Eq. 14 allows to obtain the profile $\tilde{z}(\tilde{y})$ in the dimensionless coordinates normalized by $b$. By obtaining the curvature $b$ at the apex, the droplet's profile $z(y)$ can be derived, which is sufficient for identifying the intersection points and evaluating the contact angle at each side of the 2D droplet.

In our model, we assumed the weak dependence of the surface tension on a curvature in our case. The latter was based on an estimation of the deviation of the surface tension for a bulk water from one of the nanodroplets according to Tolman approximation (Eq. 3) where $\gamma^\infty$ was taken equal to 63 mN/m, $\delta$ is the Tolman length of water equal to -0.659 Å [38] and $R_c$ is the radius of curvature in the apex of the droplet. For the droplets analysed in this work, the radius of the principal curvature ranged from 26 to 105 Å. This variation caused a change in surface tension of no more than 3%, which did not significantly impact the droplet's shape predicted by the model.

[38]According to the model, adsorption does not affect the droplet's shape, but only its height and volume. This, in turn, alters the position of the intersection points.

The model provides the advancing $\theta_a$ and receding $\theta_r$ contact angles in the ideal case, where there is a perfect sliding ("no-friction") of the liquid on the solid surface. This is obviously not physical; hence some discrepancies are expected when comparing the above model with MD simulations.

# 3. Results and discussions

The model was used to fit the droplet edge (red line in Figure 4) evaluated from MD simulations. The examples of droplet shapes fitted by the model are presented in Figure 6. The figure shows density profiles of droplets with different wettability regimes - the most hydrophobic, neutral, and the most hydrophilic - under an external force of 0.198 nN. The red line on the graph denotes the droplet's shape, indicating the region where the density of water equals half of its bulk density, while the green curve represents the prediction of the hydrostatic model.

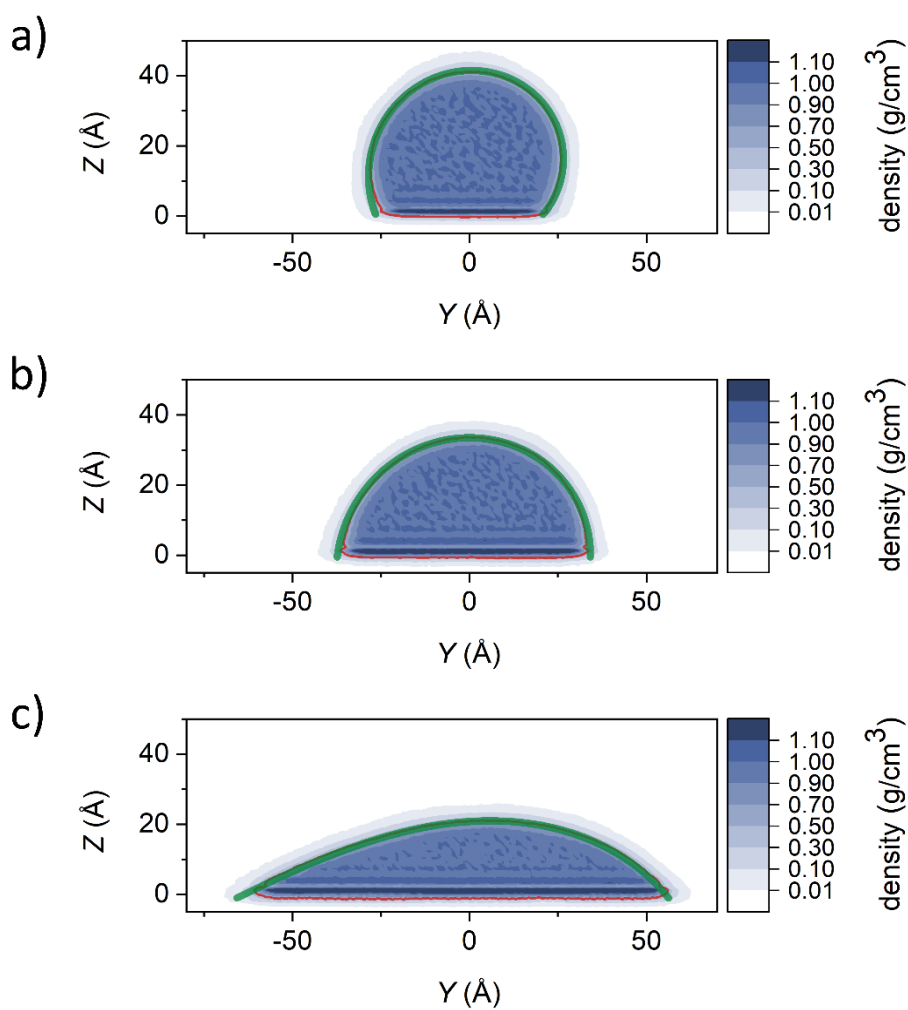

*Figure 6. Density profiles of droplet for system with: a) $\varepsilon_{Si-O}$ = 10meV, b) $\varepsilon_{Si-O}$ = 15.7meV, c) $\varepsilon_{Si-O}$ = 21 meV with exerted force $F$ = 0.198 nN; comparison with the model's prediction (green line).*

The level of convergence between model-predicted and MD-derived droplet profiles was determined by calculating the root mean square deviation (RMSD) according to the formula:

$$RMSD = \sqrt{\frac{\sum((y_{model} - y_{MD})^2 + (z_{model} - z_{MD})^2)}{n}} \tag{15}$$

where $y_{MD}$, $z_{MD}$ and $n$ are the coordinates and the total number of points defining the MD-derived droplet profile; $y_{model}$ and $z_{model}$ are the coordinates of the closest point on the model-predicted profile.

If the RMSD value was below 0.35 Å, it was assumed that a perfect fit was achieved, indicating a high degree of agreement between the two curves. A value between 0.35 and 0.45 Å was classified as a partial fit, suggesting some deviation between the curves. Finally, if the RMSD value exceeds 0.45 Å, it was considered a substantial misfit between the predicted curve and the actual droplet shape obtained from the molecular dynamics method.

Wetting angles were calculated as tangents to the model-predicted curve at the points of intersection with the substrate. The resulting wetting angles dependence on the Bond number and $\varepsilon$-parameter for Si-O interaction is presented in Figure 7.

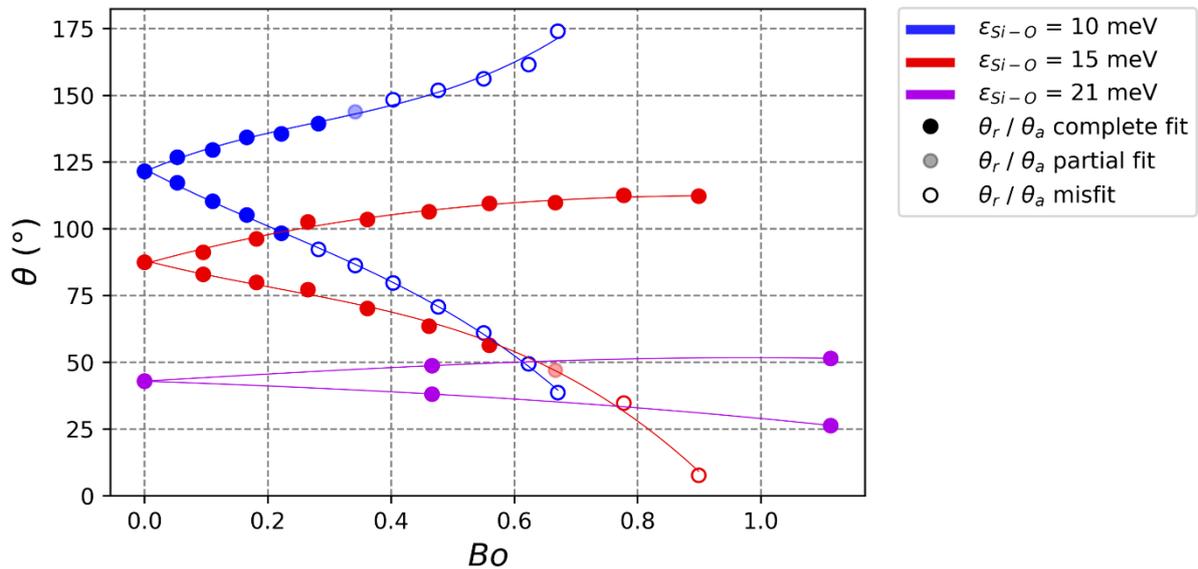

*Figure 7. Dependence of wetting angle on the Bond number for $\varepsilon_{Si-O}$ = 10, 15 and 21meV. For each $\varepsilon_{Si-O}$ value, the curve that rises represents the advancing contact angle, while the descending curve represents the receding contact angle.*

The lines displayed on the graph depict the eye-guided paths obtained using a third-degree polynomial approximation. Different symbols indicate the level of agreement of the droplet profiles. As the Bond number increases, the difference between the advancing and receding angles becomes greater. The advancing angle increases, while the receding angle decreases. It can be observed that this simplified model is efficient to predict the droplet shape when the value of $\varepsilon_{Si-O}$ is high. In fact, the contact angles $\theta_a$ and $\theta_r$ are functions of the velocity. The greater the wettability of the surface, the lower the velocity that the droplet reaches. This behavior is illustrated in Figure 3.

It can also be seen a rough linear relationship between the contact angle hysteresis $\Delta\theta = \theta_a - \theta_r$ and the Bond number defined from the component of the force tangential to the surface plane (see Figure 8). In addition, we can observe the decrease of slope of this dependence with increase of $\varepsilon_{Si-O}$. Thus, it is clear that the CAH is more pronounced for the hydrophobic case for the same Bond number. This could be attributed to the pinning of the contact line at the atomistic roughness. Specifically, as evident from Figure 10 for the hydrophobic case, the tangential component of velocity near the advancing contact line tends towards zero, while the tangential component near the receding contact line exhibits a significant magnitude, directed inversely to the external force.

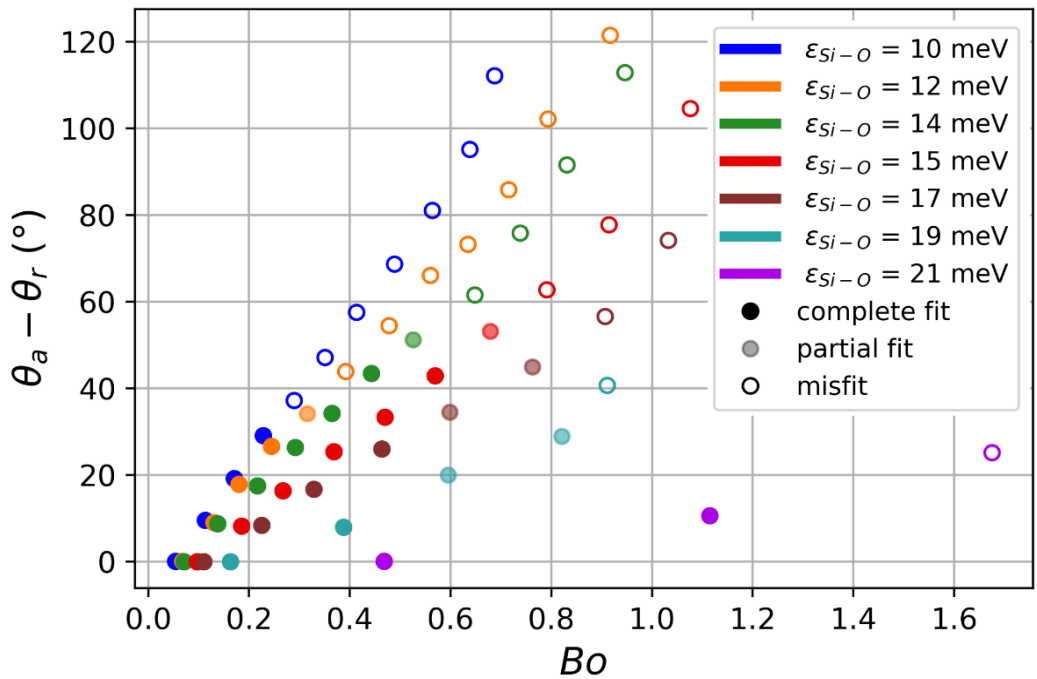

*Figure 8. Contact angle hysteresis Δθ as a function of Bo.*

Since the model was based on assumption of the stationary droplet we should estimate the frames of its applicability. This may be done with use of the capillary number ($Ca$) which defines the interplay between capillary and viscous forces:

$$Ca = \frac{v \cdot \mu}{\gamma_{sl}} \qquad (16)$$

where $v$ – velocity of a droplet, $\mu$ – shear (dynamic) viscosity (0.729 mPa·s).

Figure 9 demonstrates the strong correlation between the capillary number based on the droplet velocity and the Bond number. The cross-symbols represent the cases when the droplet is greatly elongated and a so-called "tail" is formed at the trailing edge of the moving droplet. In most of the descriptions of contact line dynamics, the value of Ca = 0.1 is used as a threshold. When the capillary number exceeds this value, droplet behaviours become more complex, and the models may need to be more accurate [57].

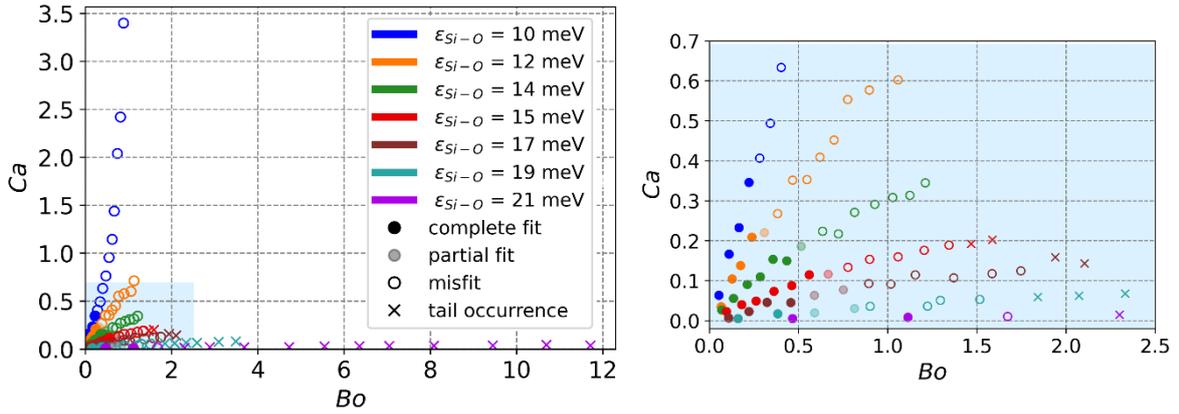

*Figure 9. Dependence of the capillary number on the Bond number.*

Figure 9 suggests that there is roughly a linear relationship between $Bo$ and $Ca$ at least for moderate values of Bo. These two dimensionless numbers are connected by the balance of forces. Three forces are acting on the droplet: the volume force $F$, the viscous drag on the solid surface and the interfacial capillary force. Based on Figure 5, the in-plane component of the volume force $F$ is:

$$F = \rho \, V_o \, g \qquad (17)$$

where $V_o$ is the volume of the droplet. The viscous drag can be determined by:

$$F_\mu = -\mu\, l\, v \tag{18}$$

where $l$ is a characteristic length. Here, the effect of viscosity spreads over the entire volume of the nanosized droplet and we can choose $l = V_o^{1/3}$. The interfacial capillary force corresponds to the net force acting on the contact line. It is related to the difference between the advancing and receding contact angles (Eq. 2) The advancing and receding angles obey $\theta_a = \theta_m + f\,\Delta\theta$ and $\theta_r = \theta_m + (1-f)\,\Delta\theta$, where $0 \leq f \leq 1$. $\theta_m$ is the static contact angle in the absence of motion of the triple contact line. According to [58],

$$F_c = -\gamma_l\, r \sin\theta_m\, \Delta\theta \tag{19}$$

The force balance implies the following relationship:

$$\frac{\rho\, g}{\gamma_{ls}} \frac{V_o}{r} - \sin\theta_m\, \Delta\theta = \frac{\mu\, v}{\gamma_{ls}} \frac{V_o^{1/3}}{r} \tag{20}$$

$$Bo\, \frac{V_o}{r\, b^2} - \sin\theta_m\, \Delta\theta = Ca\, \frac{V_o^{1/3}}{r} \tag{21}$$

Parameters $\frac{V_o}{r\, b^2}$ and $\frac{V_o^{1/3}}{r}$, related to the shape of the droplet, are of the order of 1. The factor $(\sin\theta_m)$ is only function of the interaction factor $\varepsilon_{Si-O}$. As shown in Figure 8, the contact angle hysteresis $\Delta\theta$ is approximately increasing as $f_1(\varepsilon_{Si-O}) \cdot Bo$. Finally, the following linear relationship between $Ca$ and $Bo$ is obtained:

$$Ca \approx Bo \cdot (1 - \sin\theta_m \cdot f_1) \tag{22}$$

The theoretical model proposed in section 2 relies on simplifying that there are no liquid flows inside the droplet, although such flows are present in practice. Specifically, we calculated the mass fluxes occurring inside the droplet. Figure 10 illustrates the vector field of these mass fluxes averaged over 1000 snapshots recorded during 1 ns. The contribution to the mass transfer due to the translational motion of the droplet was eliminated by subtracting the velocity of the center of mass at each time step. We observe a transition from sliding to rolling movement of the droplet as the force increases. Moreover, it becomes evident that the trends for rolling movement become more pronounced as hydrophobicity increases. Thus, we can conclude that internal fluxes are crucial in determining droplet movement.

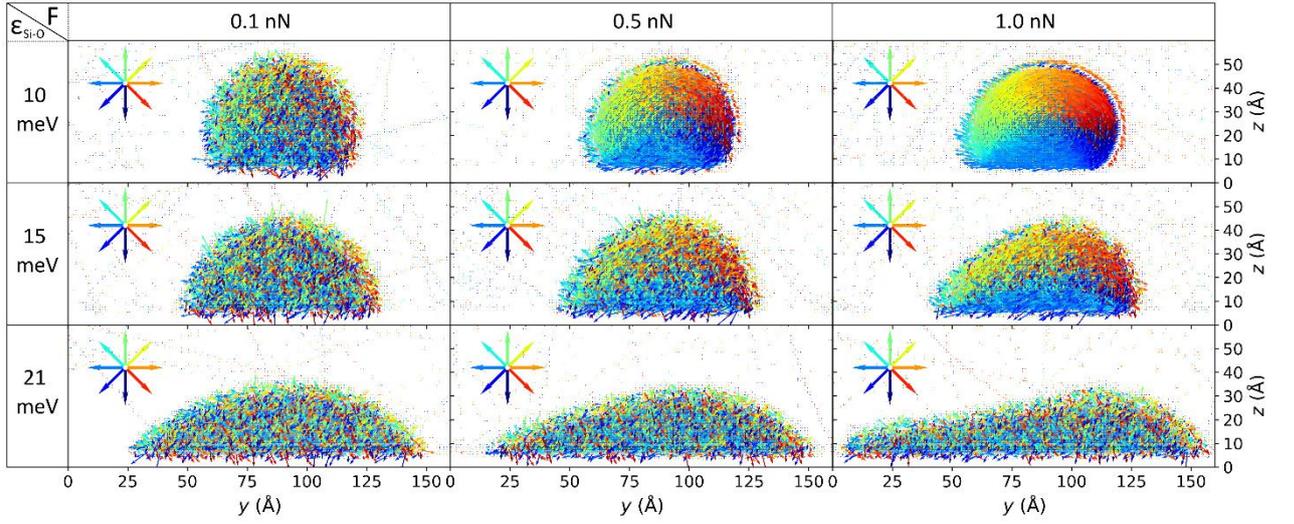

*Figure 10. Vector field of mass fluxes inside the droplets under external forces of 0.1, 0.5, and 1.0 nN for the cases of $\varepsilon_{Si-O}$: 10 meV, 15 meV, and 21 meV.*

In order to estimate this impact, we considered the circulation field of the velocity. The circulation of a velocity field inside a droplet around contour $C$ can be calculated as:

$$\Gamma = \oint_C \vec{v}\vec{dl} \qquad (23)$$

where $C$ is an isodensity curve representing the contour of the droplet. The average velocity of circulation along the contour can be obtained by dividing the circulation by the length of the contour:

$$v_{circ} = \frac{\Gamma}{L} \qquad (24)$$

In this way, it is possible to obtain an analogue of the capillary number, which will characterize the circulation of flows inside the droplet.

$$Ca^* = \frac{v_{circ} \cdot \mu}{\gamma_l} \qquad (25)$$

Its dependence on the Bond number is shown in Figure 11.

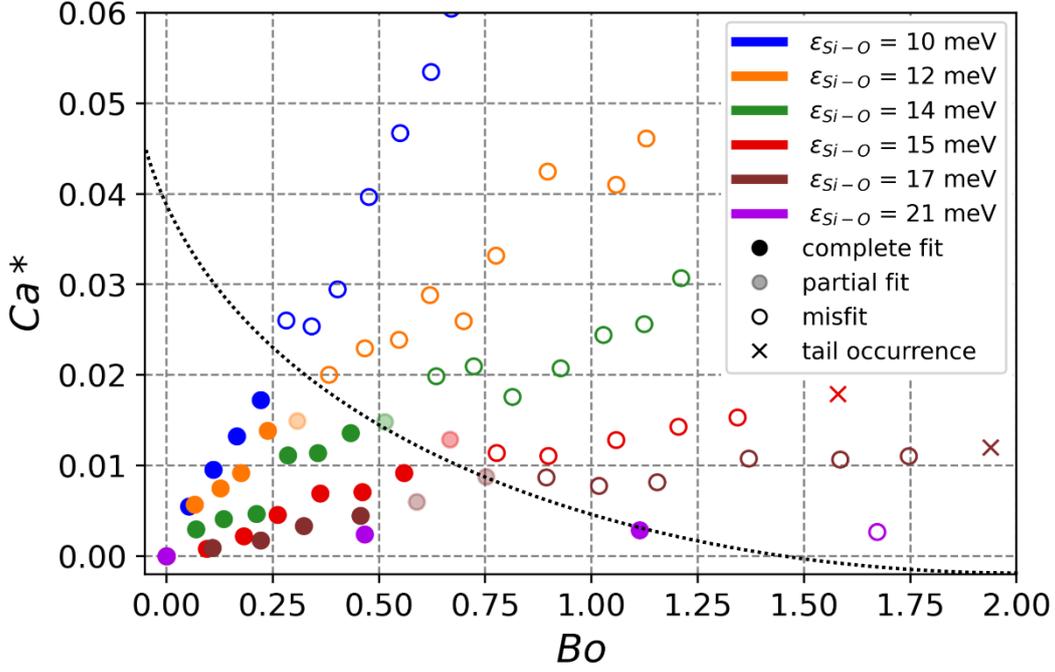

*Figure 11. Circulating capillary number for the cases of $\varepsilon_{Si-O}$: a) 10 meV (blue), b) 12 meV (orange), 14 meV (green), 15 meV (red), 21 meV (purple)*

In the case of the circulating capillary number, as in the case of the conventional one, the dependence on the Bond number remains linear. The dashed line roughly represents the area where the simplified model is applicable with a pretty good accuracy. We assume that the circulation capillary number can serve as a certain criterion for the applicability of the model in the region of low Bond numbers. Exceeding a certain threshold will indicate that the flows inside the droplet become significant making static models ineffective. In the case of high $Ca^*$ rotational movement of interface perturbates the droplet shape with respect to the one predicted by the capillary model. Thus, the rotation may impact crucially on the geometry of the interface between liquid and vapor.

In order to advance the following analysis, we correlated our results with the classical Cox-Voinov model for contact line movement [59]. In frames of this model the function $g_{CV}(\theta)$ is introduced as follows:

$$g_{CV}(\theta) = \int_0^\theta \frac{x - \sin x \cos x}{2 \sin x} dx \qquad (26)$$

here, $\theta$ refers to the receding contact angle when $Ca < 0$ and the advancing contact angle when $Ca > 0$. The corresponding equilibrium angle with no force applied is denoted as $\theta_m$.

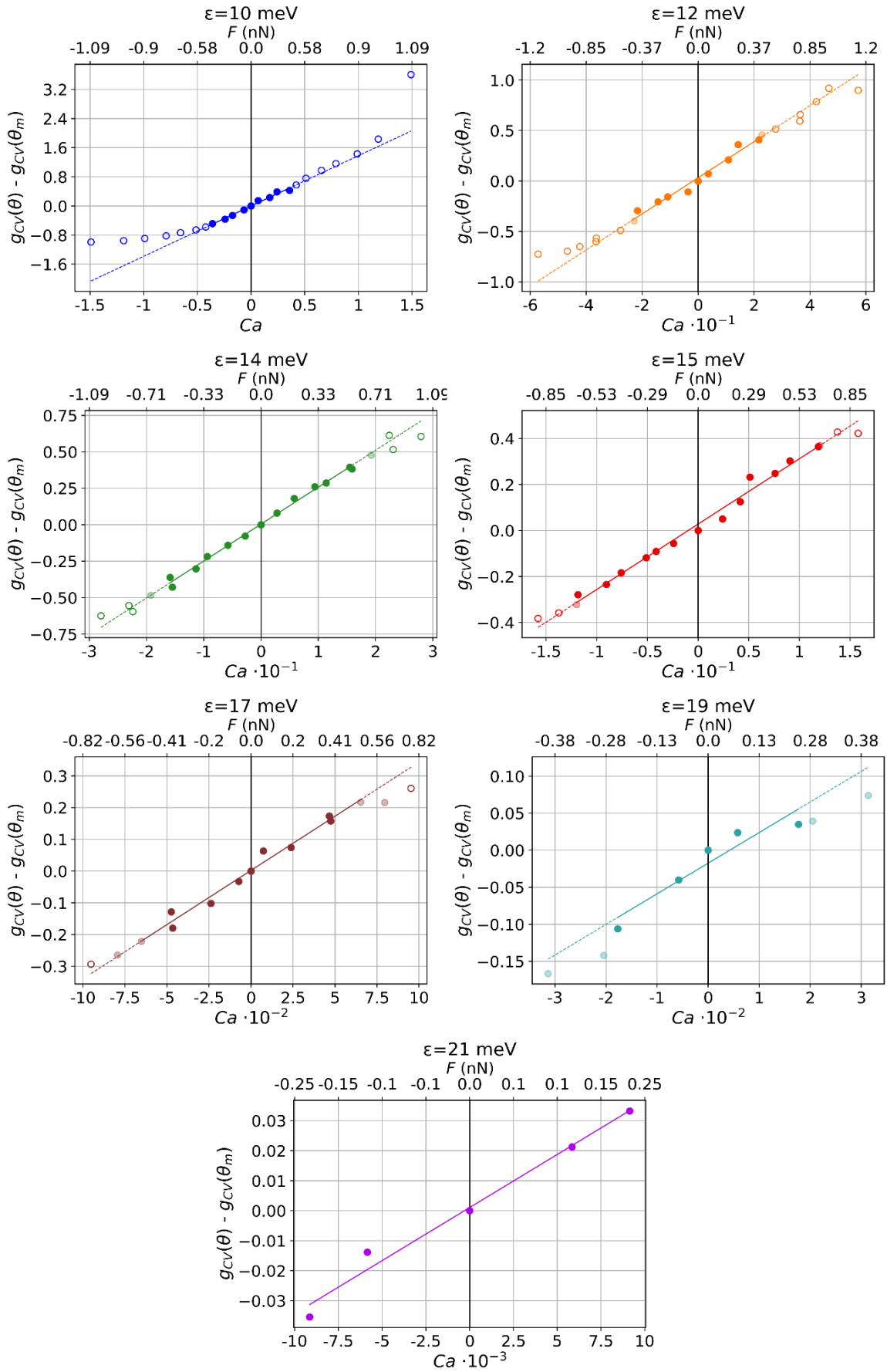

*Figure 12. Cox-Voinov model for the dynamics of wetting*

According to Cox-Voinov model the dependence of $g_{CV}(\theta)$ on capillary number should be linear. Figure 12 illustrates this dependence for all considered cases. As one can see, the deviation from linear trend for our data occurs for the points where the proposed capillary model fails to predict the droplet shape. To enhance the clarity of the simulation conditions, the upper x-axis on each graph of Figure 12 depicts the force exerted on the droplet to achieve the indicated capillary numbers.

### 3.1. Size dependence of contact angle hysteresis

It should be noted that the CAH may be dependent on the droplet size. In order to verify this, we decided to consider bigger droplet. All the size of the simulation domain are presented in Table 2. The initial droplet shape is presented in Figure 13a.

Table 2. Lengths of the simulation box, silicon substrate, and water droplet, and the number of silicon wafer and droplet atoms for the bigger droplet.

| Simulation box | | 108.6 Å×325.7 Å×200.0 Å |
|---|---|---|
| Silicon substrate | | 108.6 Å×325.7 Å×43.44 Å |
| Initial configuration of a water droplet | | 105.5 Å×62.0 Å×46.5 Å |
| Number of atoms | Silicon | 4×20×75×8 = 48 000 |
| | Water | 3×34×20×15 = 30 600 |

Figure 13 also shows the molecular snapshot (Figure 13b) and density profile (Figure 13c) for the force with magnitude equal to 0.44 nN. As one can see, the droplet shape is also well fitted with respect of the simulation model describe in Section 2.

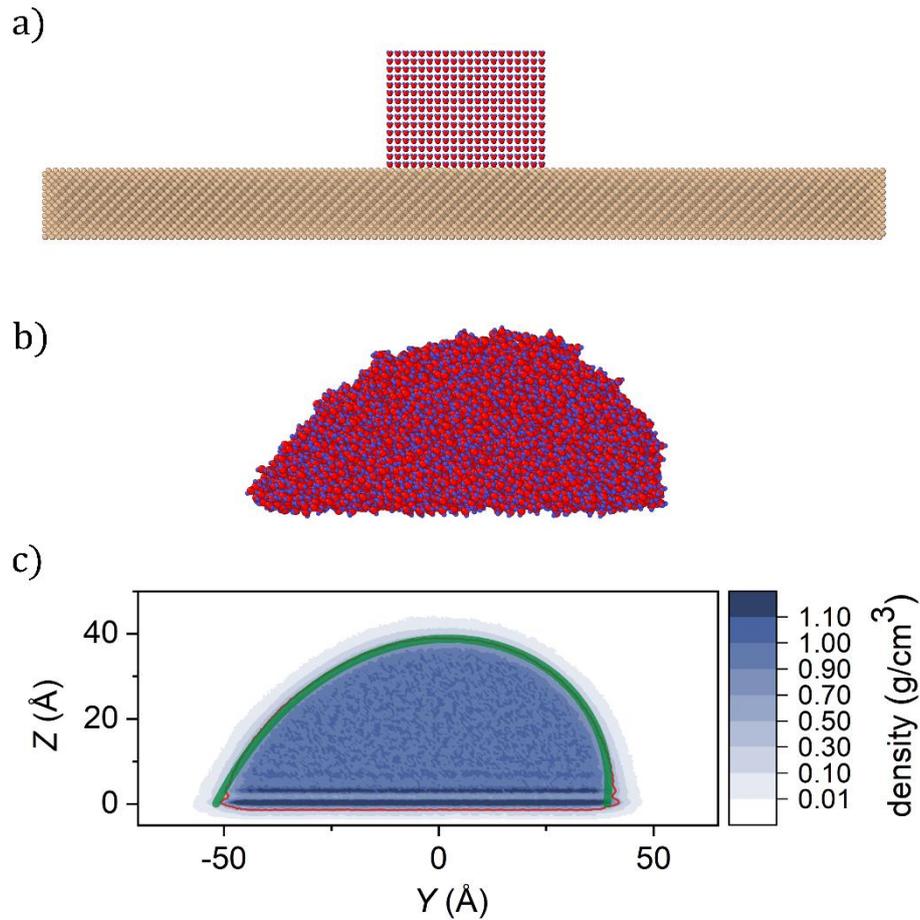

*Figure 13. The initial system (a), a snapshot of the droplet that has reached a uniform mode of motion (b), and a density profile (c), for the case of a larger droplet size, for the wetting regime $\varepsilon_{Si-O}$ = 17 meV, under the external force equal to 0.44 nN.*

The resulting dependences of the receding and advancing contact angle on Bond number are presented in Figure 14. Whenever the advancing contact angle is slightly higher for the bigger droplet, generally the size effect is unsignificant.

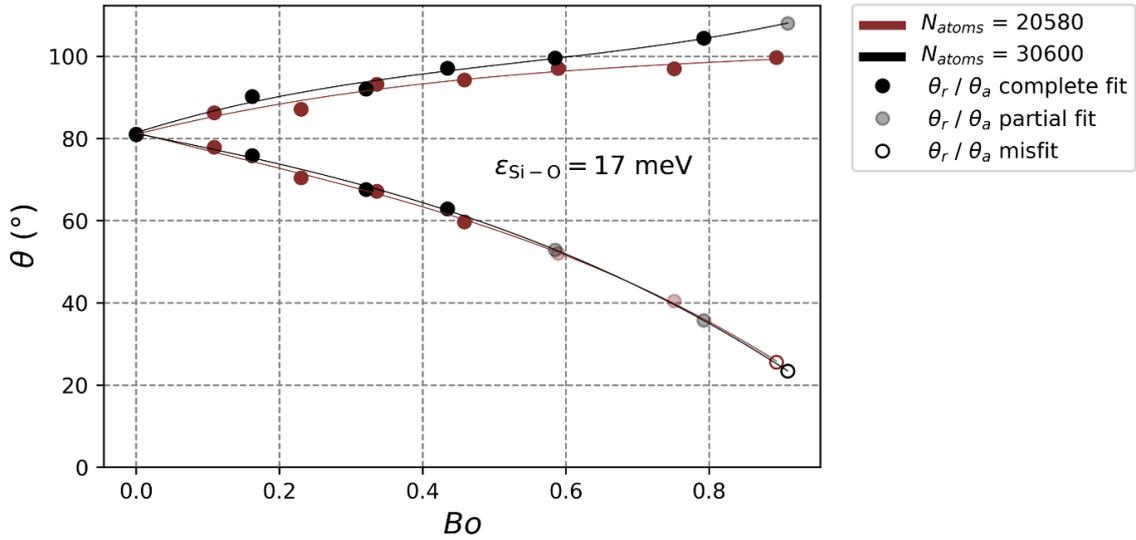

*Figure 14. Dependence of receding and advancing contact angles on the Bond number for standard (Table 1, brown line) and larger (Table 2, black line) droplets, for the wetting regime $\varepsilon_{Si-O}$ = 17 meV.*

## Conclusions

Inspired by the intricate dynamics of three-phase contact lines at the nanoscale, our research uses molecular dynamics simulations to explore water nanodroplets on solid substrates subjected to external volumetric forces. We developed an analytical capillary model that elegantly predicts the evolving shape of these nanodroplets. To validate our model, we compared its predictions with the outcomes of molecular dynamics simulations. This comparison allowed us to extract valuable insights into the wetting behavior, specifically by evaluating both the receding and advancing wetting angles.

What distinguishes our study is the emphasis on the interplay between capillary and viscous forces, which plays a pivotal role in sculpting the nanodroplet's contour under a constant external force. Our findings reveal that this dynamic interaction governs the formation of the droplet shape. Furthermore, we have demonstrated that, particularly within the realm of moderate capillary numbers, our model aligns well with the well-established Cox-Voinov model of moving contact lines, affirming the robustness and applicability of our approach.

### Declaration of interests

The authors declare that they have no known competing financial interests or personal relationships that could have appeared to influence the work reported in this paper.

## Data availability statement

The data that support the findings of this study are available from the corresponding author upon reasonable request.

## Acknowledgment

This paper presents results obtained in the frames of the project DropSurf (ANR-20-CE05-0030) and PROMENADE (ANR-23-CE50-0008-01). This work was performed using HPC resources from GENCI-TGCC and GENCI-IDRIS (A0130913052), in addition, HPC resources were partially provided by the EXPLOR center hosted by the Université de Lorraine. Thanks to "STOCK NRJ" which is co-funded by the European Union within the framework of the Program FEDER-FSE Lorraine and Massif des Vosges 2014–2020.

We appreciate Prof. Laurent Chaput's fruitful discussions.